\documentclass{aastex}
\usepackage{spr-astr-addons}
\usepackage{url}

\RequirePackage{color}

\begin{document}

\title{Luminosity-Colour Relations for Red Clump Stars}
\shorttitle{Luminosity-Colour Relations for RCs}
\shortauthors{S. Bilir, \"O. \"Onal, S. Karaali, A. Cabrera-Lavers, H. \c{C}akmak}

\author{S. Bilir \altaffilmark{1}}
\altaffiltext{1}{Istanbul University, Faculty of Science, Department 
of Astronomy and Space Sciences, 34119 University, Istanbul, Turkey\\
\email{sbilir@istanbul.edu.tr}}
\and
\author{\"O. \"Onal\altaffilmark{1}} 
\altaffiltext{1}{Istanbul University, Faculty of Science, Department 
of Astronomy and Space Sciences, 34119 University, Istanbul, Turkey\\}
\and
\author{S. Karaali\altaffilmark{1}} 
\altaffiltext{1}{Istanbul University, Faculty of Science, Department 
of Astronomy and Space Sciences, 34119 University, Istanbul, Turkey\\}
\and
\author{A. Cabrera-Lavers\altaffilmark{2,3}}
\altaffiltext{2}{Instituto de Astrof\'{\i}sica de Canarias, E-38205 La Laguna, Tenerife, Spain}
\altaffiltext{3}{Departamento de Astrof\'{\i}sica, Universidad de La Laguna, E-38205 La Laguna, Tenerife, Spain}
\and
\author{H. \c{C}akmak\altaffilmark{1}}
\altaffiltext{1}{Istanbul University, Faculty of Science, Department 
of Astronomy and Space Sciences, 34119 University, Istanbul, Turkey\\}

\begin{abstract}
We calibrated the $M_V$, $M_J$, $M_{K_s}$ and $M_g$ absolute magnitudes of red clump stars in terms of colours. $M_V$ and $M_g$ are strongly dependent on colour, while the dependence of $M_J$ and $M_{K_s}$ on colour is rather weak. The calibration of $M_V$ and $M_{K_s}$ absolute magnitudes is tested on 101 RC stars in the field SA 141. The Galactic model parameters estimated with this sample are in good agreement with earlier studies.
\end{abstract}


\section{Introduction}

Red Clump (RC) stars are core helium-burning giants. They form a prominent feature in the colour-magnitude diagrams (CMDs) of open clusters. Following the prediction of \cite{Cannon70}, it is known that they are abundant in the solar neighbourhood. In recent years much work has been devoted to studying the suitability of RC stars as a distance indicator. Their absolute magnitudes in the optical range lie from $M_V=+0.7$ mag for those of spectral type G8 III to $M_V=+1$ mag for type K2 III \citep{Keenan99}. The absolute magnitude of these stars in the $K_s$ band is $M_{K_s}=-1.61\pm0.03$ mag with negligible dependence on metallicity \citep{Alves00}, but with real dispersion. Based on observations of 14 open clusters with $-0.5<[Fe/H]\leq0$ dex and $1.58\leq t \leq 7.94$ Gyr, \citet{Grocholski02} found that for RC stars in clusters have $<M_{K_s}>=-1.61\pm0.04$ mag. 

The dependence of the $I$-band magnitude of RC stars on the metallicity and age was extensively studied in the past from an observational point of view. In the most cases the $I$ band-mean absolute magnitude is insensitive to age and metallicity. \citet{Udalski00} found that the $M_I$ of RC stars weakly depends on metallicity. \citet{Paczynski98} and \citet{Stanek98} found little or no variation in $M_I$ with colour and  metallicity. \citet{Sarajedini99} presented observations of eight open clusters, concluding that $M_I$ is less sensitive to metal abundance than $M_V$, but that the dependence on age and metallicity is still not negligible. \citet{Zhao01} and \citet{Kubiak02} confirmed the results of \citet{Udalski00}, and theoretical models from \citet{Girardi01} also show a dependence in $I$-band, predicting that an older cluster with higher metallicity has fainter RC stars. Based on the model of \citet{Girardi00}, \citet{Salaris02} stated that $M_K$ is a complicated function of metallicity and age. For age $t>1.5$ Gyr, it decreases with increasing metallicity, the opposite behaviour with respect to $M_V$ and $M_I$ absolute magnitudes. 

\citet{Pietrzynski03} have also investigated the dependence of the mean $K$, $J$, and $I$ absolute magnitudes of the RC stars on metallicity and age, as a part of their ongoing Araucaria Project to improve stellar distance indicators. They took deep near-infrared (NIR) $J$ and $K$ images of several fields in LMC, SMC, and the Carina, and Fornax dwarf galaxies and made a comparison between the extinction -corrected $K$-band RC star magnitudes and some other stellar indicators, particularly the tip of the red giant -branch magnitude, the mean RR Lyrae star $V$-band magnitude, and the mean $K$-band magnitude of Cepheid variables at a period of 10 days. This comparison strongly suggests that absolute $K$-band magnitude of the RC stars have a very weak dependence, (if any) on $[Fe/H]$ over the broad range of metallicities covered by their target galaxies. They conclude that the mean $K$-band magnitude of the RC stars is an excellent distance indicator with small (if any) population corrections over a wide range in metallicity and age. \citet{Pietrzynski10} stated that population effects strongly affect both $V$ -and $I$- band magnitude of RC stars in a complicated way. Therefore, optical $V-I$ photometry of RC stars is not an accurate method for the determination of distances in nearby galaxies, while NIR photometry is a much better way to measure distances with RC stars given its smaller sensitivity to population effects. \citet{Laney12} determined the mean $M_{K_s}$ absolute magnitude for RC stars in the solar neighbourhood to within 2 per cent ($M_{K_s}=-1.613\pm0.015$ mag) and applied their results to the estimation of the distance of LMC. A mean value for the $M_{K_s}$ absolute magnitude with weak dependence on metallicity makes it possible to use this population as a tracer of Galactic structure and interstellar extinction, as several works have fully demonstrated in the last decade \citep[see for example][and references therein]{Lopez02, Lopez04, Cabrera05, Cabrera07a, Cabrera07b, Cabrera08, Bilir12}.

In a recent work, \citet{vanHelshoecht07} used the Two Micron All Sky Survey \citep[2MASS;][]{Skrutskie06} infrared data for a sample of 24 open clusters to investigate how the $K_s$-band absolute magnitude of the red clump depends on age and metallicity. They showed that a constant value of $M_{K_s}=-1.57\pm0.05$ mag is a reasonable assumption to use in distance determinations of clusters with metallicity between -0.5 and +0.4 dex and age between 0.31 and 7.94 Gyr. The constant absolute magnitude value of RC stars was also supported with the newly reduced {\em Hipparcos} data by \citet{Groenewegen08} ($M_{K_s}=-1.54\pm0.04$ mag).  

In this paper, we will contribute to the discussion by a different approach. We aim to calibrate the absolute magnitudes of RC stars in four bands, i.e. $V$, $J$, $K_s$ and $g$, as a function of colour with field stars taken from different photometric surveys. The data are given in Section 2. The location of the RC stars in the Hertzsprung-Russell (H-R) diagram are demonstrated in Section 3. The luminosity-colour relations in optical and near-infrared are given in Sections 4. We tested the absolute magnitudes derived by the procedure in our study in Section 5, and a summary and conclusion is presented in Section 6. 

\section{The data}
We used three sets of data. The $BVI$ data were taken from the {\em Hipparcos} catalogue \citep{vanLeeuwen07} which also provides trigonometric parallaxes. In order to obtain reliable absolute magnitude and distances, only the stars with relative parallax error ($\sigma_\pi/\pi$) is less than or equals to 0.1 are selected. Thus, 32 144 stars are included into the sample with $BVI$ magnitudes. The second set of data consists of $JHK_s$ magnitudes. Within this sample only 32 072 stars were detected in the 2MASS catalogue \citep{Cutri03}. The {\em Hipparcos} stars were not observed in the Sloan Digital Sky Survey \citep[SDSS;][]{York00}. Hence, the $gri$ magnitudes of the sample stars were evaluated by the transformation equations of \citet{Yaz10}. Thus, the resulting catalogue consisted of three different photometries, i.e. Johnson-Cousins ($BVI$), 2MASS ($JHK_s$), and SDSS ($gri$) of all the RC stars within the Solar neighbourhood.

\subsection{Lutz-Kelker Correction}
The observed trigonometric parallaxes are biased because the volume of space per unit of parallax is not constant. The pioneers of this topic are \citet{Trumpler53}. However, \citet{Lutz73} were the first who quantified the bias. Other studies followed the work of \citet{Lutz73}. \citet{Smith87} claimed that \citet{Lutz73}'s methodology seemed to be that of Bayesian statistics. The Luzt-Kelker (LK) bias can be explained as follows. Let $\pi$ and $\sigma_{\pi}$ be the parallax and its error of a star. Then, one can define a distance $d$ with lower and upper limits, i.e. $d_{\pi+\sigma_{\pi}}$ and $d_{\pi-\sigma_{\pi}}$. Stars in a given volume can scatter to the distance $d$. Since the number of stars in the distance interval ($d$, $d_{\pi-\sigma_{\pi}}$] are more than the ones in the distance interval [$d_{\pi+\sigma_{\pi}}$, $d$], more stars from the interval ($d$, $d_{\pi-\sigma_{\pi}}$] will scatter to distance $d$ than the ones in [$d_{\pi+\sigma_{\pi}}$, $d$]. The result of this effect is that measured parallaxes cause smaller distance estimations. We used the following equation of \citet{Smith87} to correct the observed {\em Hipparcos} parallaxes \citep{vanLeeuwen07},

\begin{equation}
\pi_{0}=\pi(\frac{1}{2}+\frac{1}{2}\sqrt{1-16(\sigma_{\pi}/\pi)^{2}}~),
\end{equation}
where $\pi$ and $\pi_{0}$ are the observed and corrected parallaxes, respectively, and $\sigma_{\pi}$ denotes the error of the observed parallax. According to \citet{Lutz73} relative parallax errors, $\sigma_\pi/\pi$, larger than 0.17 are not reliable.

\subsection{Reddening}

The $E(B-V)$ colour excess of stars have been evaluated in two steps. First, we used the maps of \citet{Schlegel98} and evaluated a $E_{\infty}(B-V)$ colour excess for each star. Then, we reduced them using the following procedure \citep{Bahcall80}:
\begin{equation}
A_{d}(b)=A_{\infty}(b)\Biggl[1-\exp\Biggl(\frac{-\mid d \times \sin(b)\mid}{H}\Biggr)\Biggr].
\end{equation}
Here, $b$ and $d$ are the Galactic latitude and distance to the star, respectively. $H$ is the scale height for the interstellar dust which is adopted as 125 pc \citep{Marshall06}. $A_{\infty}(b)$ and $A_{d}(b)$ are the total absorptions for the model and for the distance to the star, respectively. $A_{\infty}(b)$ can be evaluated by means of the following equation:

\begin{equation}
A_{\infty}(b)=3.1\times E_{\infty}(B-V).
\end{equation}
$E_{\infty}(B-V)$ is the colour excess for the model taken from the \citet{Schlegel98}. Then, $E_{d}(B-V)$, i.e. the colour excess for the corresponding star at the distance $d$, can be evaluated via equation,

\begin{equation}
E_{d}(B-V)=A_{d}(b)~/~3.1.
\end{equation}
We have omitted the indices $\infty$ and $d$ from the colour excess $E(B-V)$ in the equations. However, we use the terms ``model'' for the colour excess of \citet{Schlegel98} and ``reduced'' for the colour excess corresponding to distance $d$. The total absorption $A_{d}$ used in the section and classical total absorption $A_{V}$ have the same meaning. 

\begin{figure}
\begin{center}
\includegraphics[scale=0.38, angle=0]{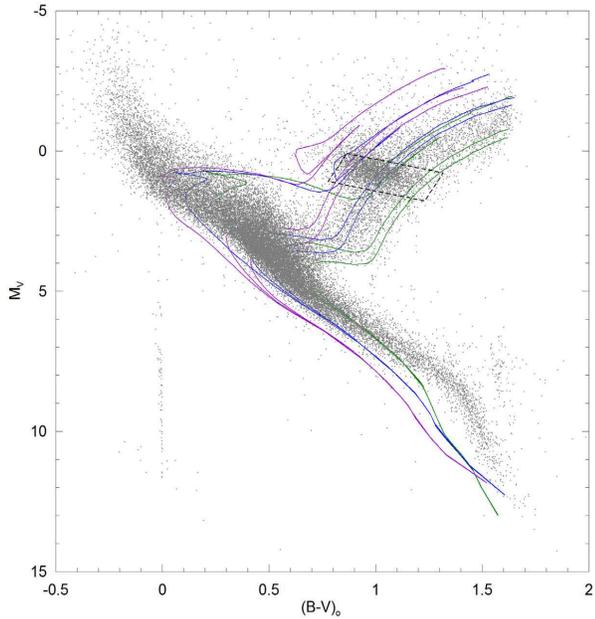}
\caption[] {The $M_V-(B-V)_0$ diagram for 32 144 stars, taken from the {\em Hipparcos} catalogue, fitted to the Padova isochrones with metallicities $[M/H]=0$, -0.5, and -1 dex, and  ages $t=1$, 5, and 10 Gyr. The dotted lines show the location of RC stars.}
\end{center}
\end{figure}

We de-reddened the colours and magnitudes using the $E(B-V)$ colour excesses of the stars evaluated from the procedures explained above and the following equations from \citet{Fan99} and \citet{Fiorucci03} for $V-I$ colour and for the 2MASS photometry. 

\begin{eqnarray}
V_{o}=V-3.1\times E(B-V),\nonumber \\
(B-V)_{o}= (B-V)-E(B-V),\nonumber \\    
(V-I)_{o}= (V-I)-1.250\times E(B-V),\\ \nonumber 
J_{o}=J-0.887\times E(B-V), \nonumber \\
(J-H)_{o}= (J-H)-0.322\times E(B-V),\nonumber \\
(H-K_{s})_{o}=(H-K_{s})-0.183\times E(B-V).\nonumber
\end{eqnarray}

\section{Hertzsprung-Russell Diagram of RC Stars}
We evaluated the $V$-band absolute magnitudes of 32 144 {\em Hipparcos} stars with relative parallax errors less than 0.1 using the following formula and plotted them onto the H-R diagram (Fig. 1) to identify the location of the RC stars:

\begin{equation}
M_V=V_0-5\log \frac{1000}{\pi (mas)}-5.
\end{equation}
The sample stars were fitted to Padova isochrones \citep{Marigo08} with metallicities $[M/H]=0$, $-0.5$, and $-1$ dex and ages $t$=1, 5, and 10 Gyr. The relatively high condensed region on the evolved segments of the isochrones corresponds to the location of the RC. The borders of the RC stars in the vertical direction have been fixed by using the constraint of \citet{Puzeras10}, i.e. $2.1\leq \log g\leq2.7$, where $g$ denotes surface gravity of the RC stars. These borders comprise the region in which RC stars identified in the literature. However, for the borders in the horizontal direction we used their highly concentrated location, i.e. $0.7\leq(B-V)_0\leq1.3$. The number of the RC stars defined in this way are 2576. We separated them into three categories with respect to their Galactic latitudes, i.e. $|b|\leq 30^{o}$, $30^{o}<|b|\leq60^{o}$, and $|b|>60^{o}$. Their median $E(B-V)$ colour excesses are 0.035, 0.020, and 0.012 mag, respectively (Fig. 2). These stars will be used in the following sections to derive colour dependent $M_V$ absolute magnitude calibrations.

The distances of 2576 RC stars evaluated by the combination of their absolute and de-reddened apparent $V_0$ magnitudes show almost a symmetrical distribution (Fig. 3). The median distance is 140 pc. Distribution of stars in $X-Y$ and $X-Z$ planes is given in Fig. 4, where $X$, $Y$, and $Z$ are the heliocentric rectangular coordinates. The distribution of stars in the medians of $X$, $Y$, and $Z$, i.e. +3, -1, and -5 pc, respectively show that the distribution of RC stars in the solar neighbourhood is almost symmetrical. 

\begin{figure}
\begin{center}
\includegraphics[scale=0.4, angle=0]{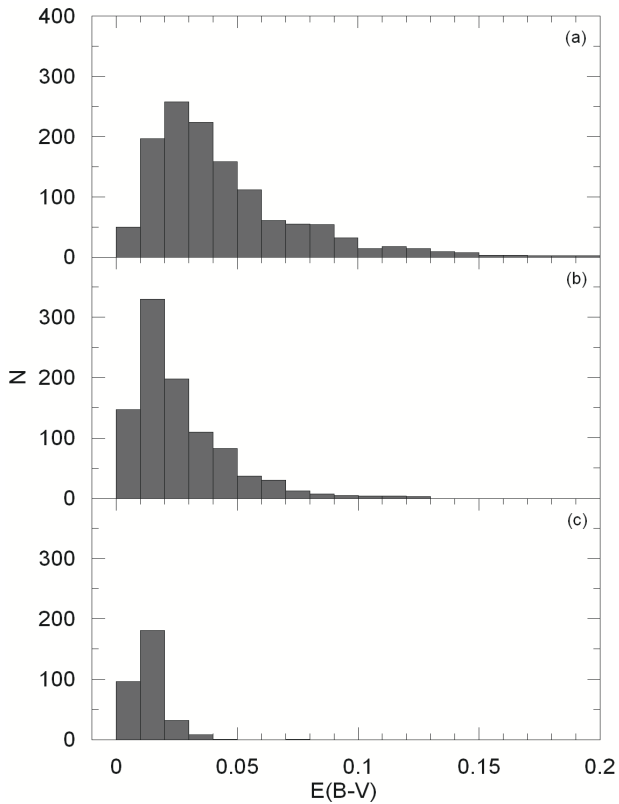}
\caption[] {Distribution of the $E(B-V)$ colour excesses of 2576 RC stars for three Galactic latitudes, $|b|\leq 30^{o}$ (a), $30^{o}<|b|\leq 60^{o}$ (b), and $|b|>60^{o}$ (c).}
\end{center}
\end{figure}

\begin{figure}
\begin{center}
\includegraphics[scale=0.4, angle=0]{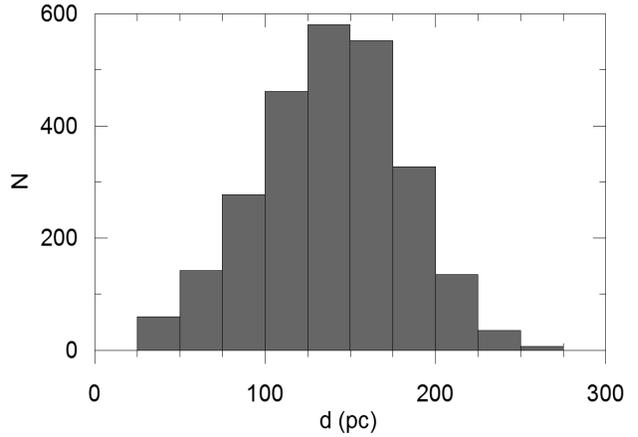}
\caption[] {Distance distribution of 2576 RC stars. The median of the distances is $d=140$ pc.}
\end{center}
\end{figure}

\begin{figure}
\begin{center}
\includegraphics[scale=0.45, angle=0]{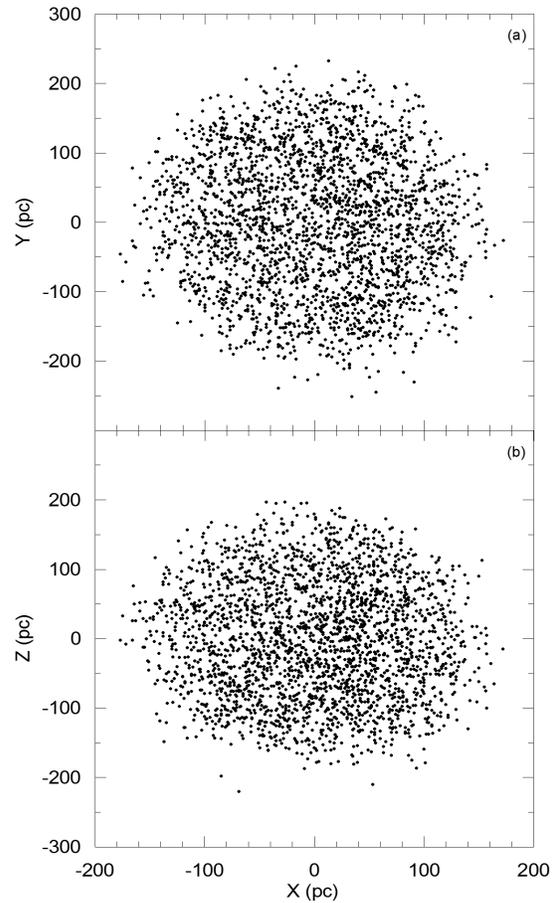}
\caption[] {Distributions of 2576 RC stars in the $X-Y$ and $X-Z$ planes.}
\end{center}
\end{figure}

The errors of magnitudes in 2MASS photometry are larger than the optical ones, $B$, $V$, and $I$. We selected the best quality $J$, $H$, and $K_s$ magnitudes by applying the constraint ``AAA'' to avoid the large errors. We found 25401 stars using this criteria. Then, we evaluated the $M_J$ and $M_{K_s}$ absolute magnitudes by means of the procedure applied to $M_V$ absolute magnitudes, and plotted them onto the H-R diagram, $M_J-(J-H)_o$ and $M_{K_s}-(J-K_s)_o$, respectively, to identify the location of the RC stars (Fig. 5 and Fig. 6) with $J$ and $K_s$ bands. As in $M_V$, the sample stars were fitted to the Padova isochrones \citep{Marigo08} mentioned above. The number of RC stars in Fig. 5 and Fig. 6 ($N=499$) are less than the ones in Fig. 1 ($N=2576$), due to the constraint ``AAA'' used to ensure the best photometric quality. 
        
\begin{figure}
\begin{center}
\includegraphics[scale=0.34, angle=0]{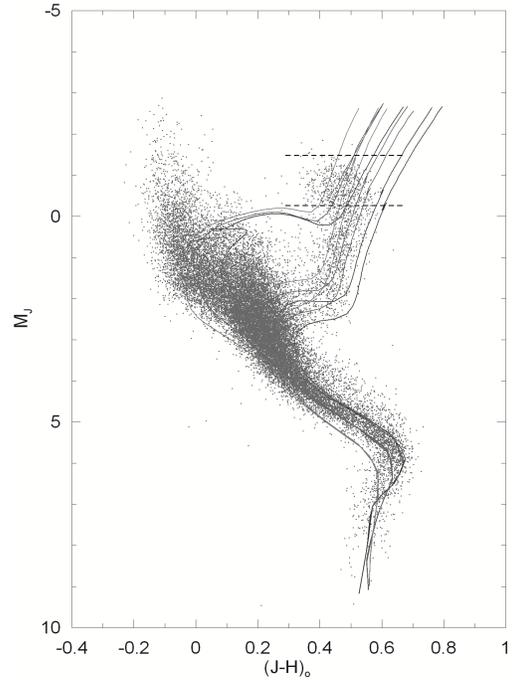}
\caption[] {The $M_J-(J-H)_0$ diagram for 25 104 stars, taken from the {\em Hipparcos} catalogue, fitted to the Padova isochrones with metallicities $[M/H]=0$, $-0.5$, and $-1$ dex, and ages $t=1$, 5, and 10 Gyr. The dotted lines show the location of the RC stars.}
\end{center}
\end{figure}

The errors of the photometric data are given in Fig. 7. One can see that the errors for the optical colour and magnitudes are smaller than the corresponding NIR ones. Also, the errors for bright magnitudes are lower than the faint ones, as expected, and the errors for $(B-V)_o$ colours are rather smaller than the ones for $(V-I)_o$ colours.

\begin{figure}
\begin{center}
\includegraphics[scale=0.34, angle=0]{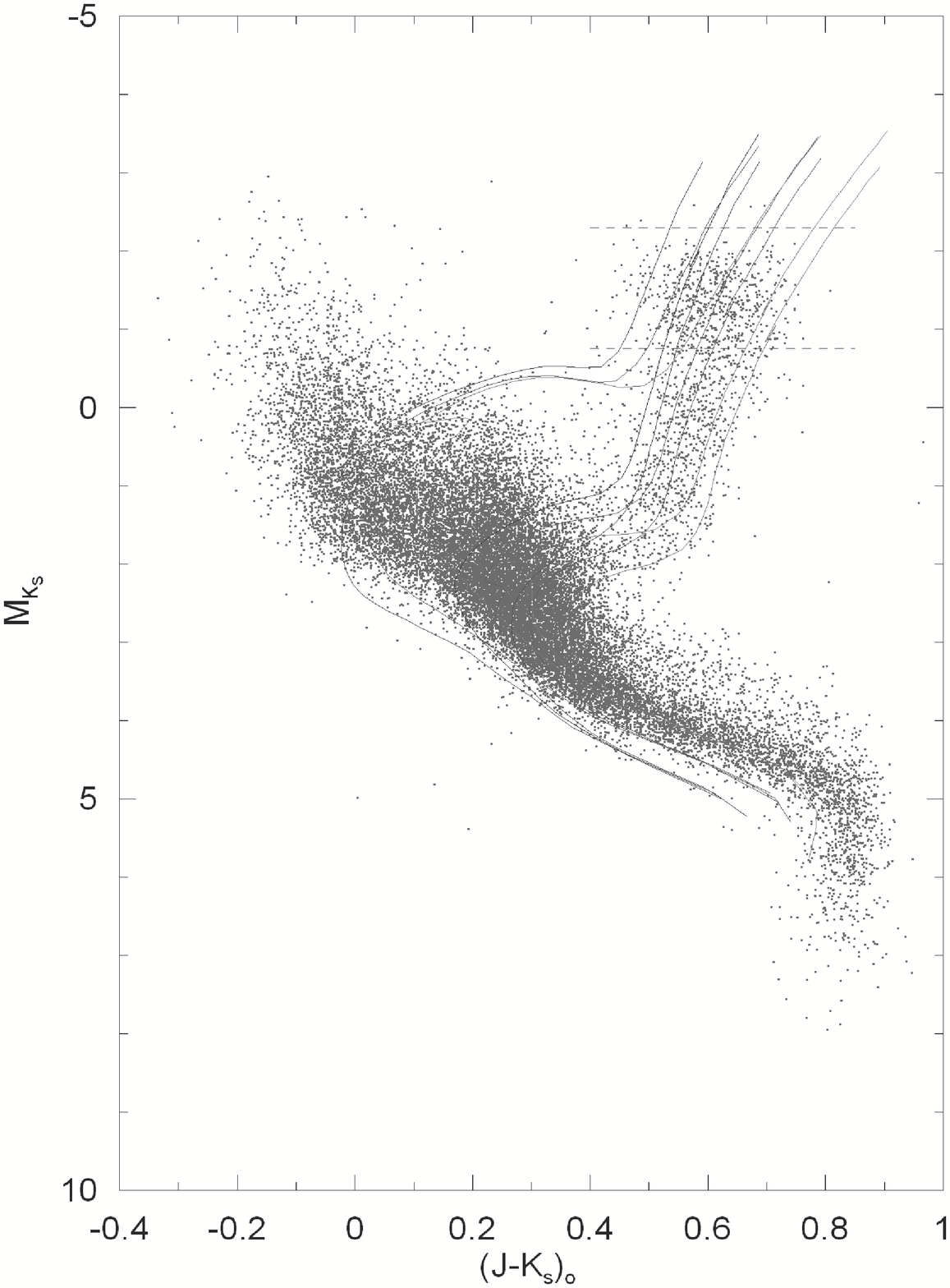}
\caption[] {The $M_{K_s}-(J-K_s)_o$ diagram for 25 104 stars, taken from the {\em Hipparcos} catalogue, fitted to the Padova isochrones with metallicities $[M/H]=0$, $-0.5$, and $-1$ dex, and ages $t=1$, 5, and 10 Gyr. The dotted lines show the location of the RC stars.}
\end{center}
\end{figure}

\begin{figure}
\begin{center}
\includegraphics[scale=0.4, angle=0]{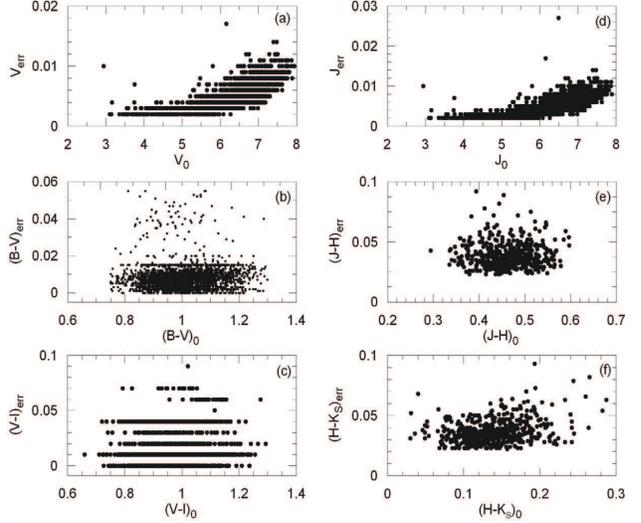}
\caption[] {Photometric errors for optical (a-c) and NIR (d-f) spectral regions.}
\end{center}
\end{figure}

We do not plot any H-R diagram for the SDSS ($gri$) data, because they were provided by the transformation equations, as mentioned above. The equation used to evaluate the $M_g$ absolute magnitude is given in the following \citep{Yaz10}:

\begin{eqnarray}
M_g-M_J=2.923(J-H)_{0}+3.031(H-K_{s})_{0}+0.329.
\end{eqnarray}

\section{Luminosity-Colour Relations for RC Stars}

We evaluated two different sets of colours for each star sample mentioned above, i.e. $(B-V)_o$ and $(V-I)_o$; $(J-H)_o$ and $(H-K_s)_o$; $(g-r)_o$ and $(r-i)_o$ for Johnson-Cousins, 2MASS and SDSS photometries, respectively, and combined them with the corresponding absolute magnitudes ($M_V$, $M_J$, $M_{K_s}$, and $M_g$) evaluated by means of their apparent magnitudes and {\em Hipparcos} parallaxes. We separated the $(B-V)_o$ and $(V-I)_o$ colours into 11 bins, whereas only six bins could be provided for the $(J-H)_o$, $(H-K_s)_o$, $(g-r)_o$ and $(r-i)_o$ colours due to smaller number of stars in 2MASS and SDSS photometries. Table 1 gives the mean colours and absolute magnitudes of the bins in question. Then, we adapted the colours and absolute magnitudes in Table 1 to the following equations and obtained colour dependent absolute magnitude equations for the RC stars by regression analysis.

\begin{eqnarray}
M_V=a_1(B-V)_{0}+b_1(V-I)_{0}+c_1,\nonumber \\
M_g=a_2(g-r)_{0}+b_2(r-i)_{0}+c_2,\\ \nonumber
M_J=a_3(J-H)_{0}+b_3(H-K_{s})_{0}+c_3,\nonumber \\
M_{K_s}=a_4(J-H)_{0}+b_4(H-K_{s})_{0}+c_4.\nonumber
\end{eqnarray}
The numerical values of the coefficients $a_i$, $b_i$, and $c_i$ ($i=1$, 2, 3, 4) estimated by means of regression analysis are given in Table 2. 

\begin{table*}
\setlength{\tabcolsep}{5pt} 
\centering
\caption{Distribution of colours and absolute magnitudes of the RC stars in 11 bins for the $BVI$ photometry, and six bins for the $JHK_s$ and $gri$ photometries.}
\begin{tabular}{cccccccccc}
\hline
\multicolumn{3}{c}{Johnson-Cousins} & \multicolumn{4}{c}{2MASS}     & \multicolumn{3}{c}{SDSS}\\
$(B-V)_o$ & $(V-I)_o$ & $M_V$   & $(J-H)_o$ & $(H-K_s)_o$ & $M_J$    & $M_{K_s}$   & $(g-r)_o$ & $(r-i)_o$ & $M_g$ \\
\hline
    0.777 & 0.777 & 0.502 & 0.339 & 0.146 & -0.937 & -1.386 & 0.571 & 0.179 & 0.825 \\
    0.830 & 0.820 & 0.574 & 0.378 & 0.142 & -0.910 & -1.400 & 0.613 & 0.196 & 0.954 \\
    0.877 & 0.863 & 0.639 & 0.431 & 0.137 & -0.873 & -1.419 & 0.671 & 0.219 & 1.131 \\
    0.927 & 0.904 & 0.709 & 0.476 & 0.132 & -0.841 & -1.436 & 0.719 & 0.239 & 1.279 \\
    0.975 & 0.945 & 0.775 & 0.522 & 0.130 & -0.809 & -1.452 & 0.772 & 0.261 & 1.440 \\
    1.024 & 0.986 & 0.843 & 0.564 & 0.126 & -0.779 & -1.467 & 0.817 & 0.279 & 1.580 \\
    1.074 & 1.028 & 0.913 &       &       &        &        &       &       &  \\
    1.122 & 1.096 & 0.979 &       &       &        &        &       &       &  \\
    1.172 & 1.144 & 1.048 &       &       &        &        &       &       &  \\
    1.222 & 1.201 & 1.118 &       &       &        &        &       &       &  \\
    1.271 & 1.227 & 1.186 &       &       &        &        &       &       &  \\
\hline
    \end{tabular}
\end{table*}

Table 1 gives the indication of an absolute magnitude gradient with respect to the colours for $M_V$, $M_g$ and $M_J$, whereas it is almost zero for $M_{K_s}$. We plotted the absolute magnitudes estimated for the bins of different colours in Table 1 versus the corresponding colour with (absolutely) larger coefficient, i.e. $(B-V)_o$, $(J-H)_o$, and $(g-r)_o$ to treat the problem schematically. Fig. 8 confirms the result obtained via Table 1, i.e. there is a linear variation of $M_V$, $M_g$ and $M_J$, whereas the locus of $M_{K_s}$ is a horizontal line. 

We evaluated the differences between the absolute magnitudes estimated using the corresponding equation in (8) as the combination of apparent magnitudes and {\em Hipparcos} parallaxes of the stars, and plotted them in Fig. 9. The original values of the absolute magnitudes are those evaluated by the combination of apparent magnitudes and {\em Hipparcos} parallaxes of the stars. The standard deviations for the optical and NIR regions are $\sigma=0.23$ and $\sigma=0.28$ mag, respectively. The dotted lines in the figure correspond to the residuals of $\pm1\sigma$.

\begin{figure}
\begin{center}
\includegraphics[scale=0.50, angle=0]{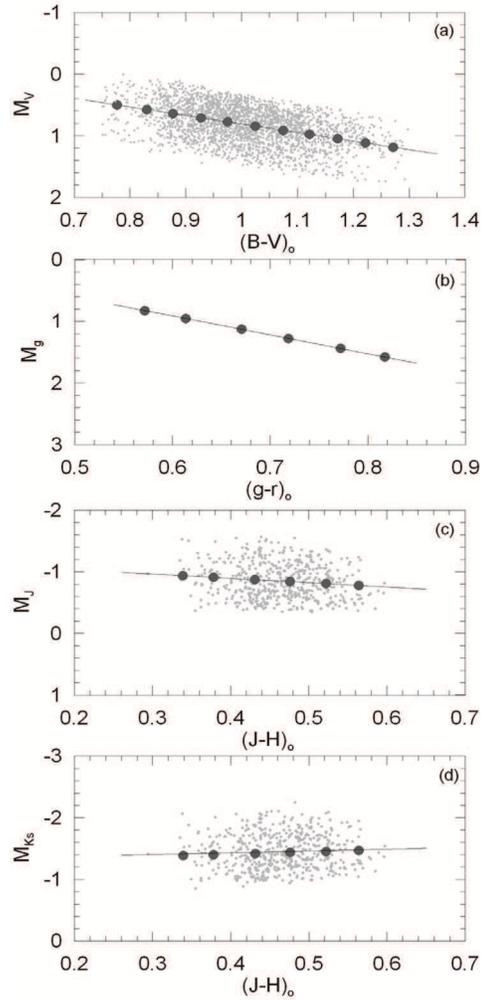}
\caption[] {Four absolute magnitude-colour diagrams for the RC stars: (a) $M_V-(B-V)_0$, (b) $M_g-(g-r)_0$, (c) $M_J-(J-H)_0$, and $M_{K_s}-(J-H)_0$.}
\end{center}
\end{figure}

\begin{figure}
\begin{center}
\includegraphics[scale=0.38, angle=0]{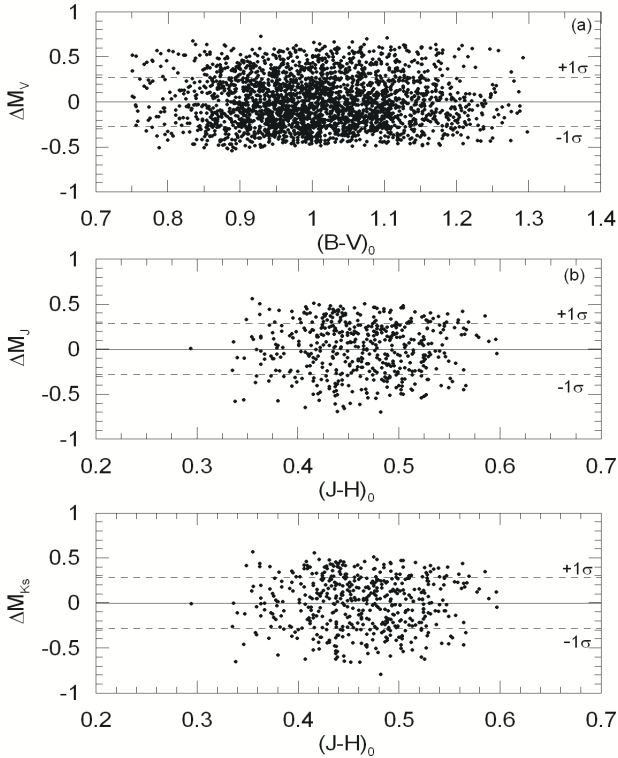}
\caption[] {Absolute magnitude residuals for $M_V$, $M_J$, and $M_{K_s}$. Dotted horizontal lines indicate   $\pm1\sigma$ residuals.}
\end{center}
\end{figure}

\section{Test of Absolute Magnitudes of RC Stars}
We tested the calibrations of $M_V$ and $M_{K_s}$ absolute magnitudes on 101 RC stars in the field SA 141. We estimated the Galactic model parameters with CCD $UBVRI$ photometric data for these RC stars and compared them with the ones appeared in the literature. The procedure is given in the following.

\subsection{Identification of the RC stars in SA 141}
\citet{Siegel09} determined the CCD $UBVRI$ magnitudes of 1299 stars in 1.2 square-degrees in a field in the direction of SA 141 ($l=246^{o}.33$, $b=-85^{o}.83$). We used the $B$ and $V$ magnitudes of these stars, identified the RC stars in this field and estimated the Galactic model parameters by using the space densities evaluated for the RC sample as explained in the following. We de-reddened the $B$ and $V$ magnitudes by using the interstellar extinction maps of \citet{Schlegel98} and the canonical procedure, as explained Section 2.2. The apparent $V$ magnitude of the stars lie in the interval of $12<V<22$ mag and their colour excesses are rather small, i.e. $0.010<E(B-V)<0.025$ mag. We added 49 stars with $JHK_s$ magnitudes to this sample which were provided from 2MASS All Sky Catalog of point sources \citep{Cutri03} and which could not be observed with $UBVRI$ photometry due to the saturation of the CCDs. Thus, the number of stars in the field increased to 1348. The colour excess for the bright stars adopted as the mean of the colour excesses of 1299 stars, $E(B-V)=0.016$ mag. 

\begin{figure}
\begin{center}
\includegraphics[scale=0.45, angle=0]{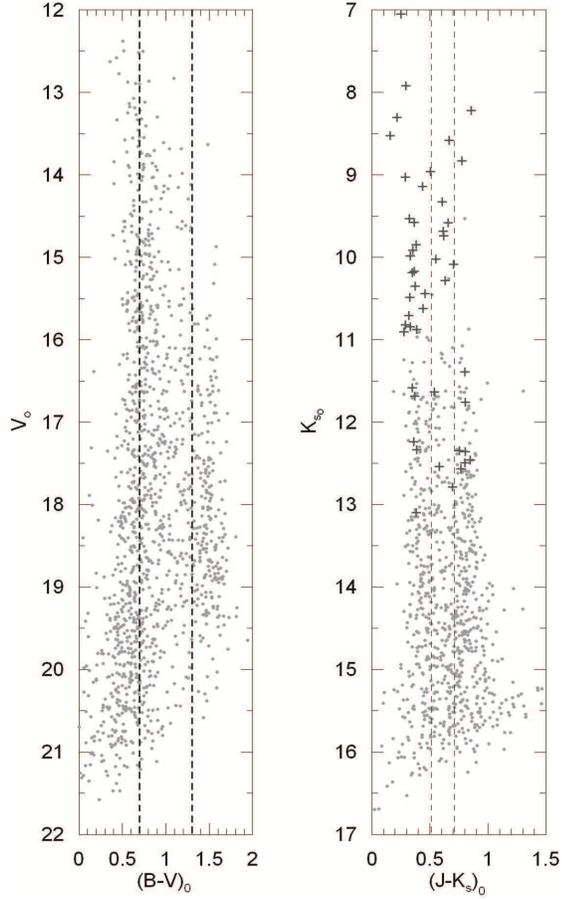}
\caption[] {$V_o-(B-V)_o$ and $K_{s_{o}}-(J-K_s)_o$ apparent magnitude colour diagrams for the stars observed in the field SA 141. The symbol ({+}) denotes the bright stars which could not be observed with $BVI$ photometry due to saturation. The vertical dashed lines indicate the lower and upper limiting colours of the RC stars.}
\end{center}
\end{figure}

\begin{table}
\setlength{\tabcolsep}{4pt} 
\centering
{\scriptsize
\caption{The numerical values of the coefficients $a_i$, $b_i$, and $c_i$ ($i=1$, 2, 3, 4) in the Eq. (8).}
\begin{tabular}{ccccc}
\hline  
   $i\rightarrow$       & (1)  & (2)  & (3) & (4)\\
\hline
      & $M_V$   & $M_g$  & $M_J$ & $M_{K_s}$\\
\hline
    $a_i$ &  1.398$\pm$0.010 &  3.152$\pm$0.075 &  0.706$\pm$0.017 & -0.337$\pm$0.013\\
    $b_i$ & -0.011$\pm$0.005 & -0.213$\pm$0.138 &  0.039$\pm$0.019 &  0.275$\pm$0.143\\
    $c_i$ & -0.577$\pm$0.001 & -0.937$\pm$0.075 & -1.182$\pm$0.034 & -1.312$\pm$0.025\\
    $s$   & 0.0003           &  0.0009          &  0.0003          &  0.0002 \\
\hline  
    \end{tabular}%
}
\end{table}%

The $V_0-(B-V)_0$ colour-magnitude diagram (CMD) of 1299 stars observed by \citet{Siegel09} and the $K_{s_o}-(J-K_s)_o$ CMD of 859 stars provided by the 2MASS catalogue \citep{Cutri03} are given in Fig. 10. Bright stars which could not be observed with $UBVRI$ photometry are marked with the symbol ($+$). The colour and magnitude errors for the $JHK_s$ photometry are larger than the ones for $UBVRI$ photometry (Fig. 11). However, 49 stars used in our study are bright ones with best quality labeled as ``AAA''. Hence, the large errors in question will not affect our results. The limiting magnitudes of completeness for $V_0$ and $K_{s_0}$ are 18.5 and 14.5 mag, respectively (Fig. 12). 

\begin{figure}
\begin{center}
\includegraphics[scale=0.40, angle=0]{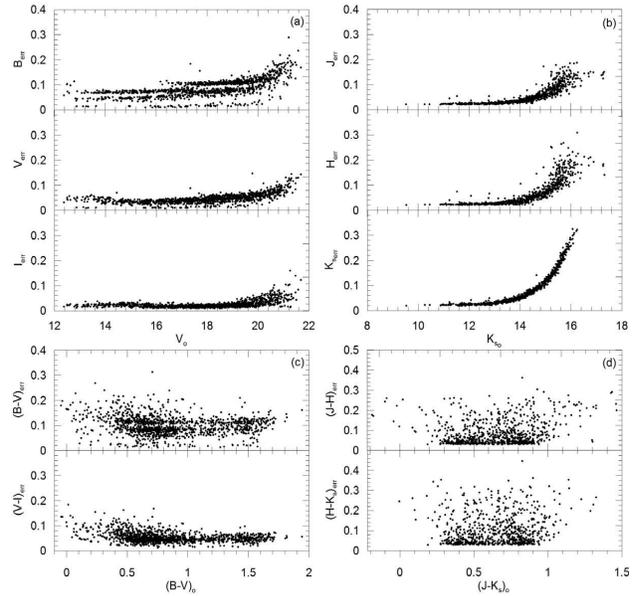}
\caption[] {Optical (a and c) and NIR (b and d) magnitude and colour errors for stars observed in SA 141 star field.}
\end{center}
\end{figure}

\begin{figure}
\begin{center}
\includegraphics[scale=0.4, angle=0]{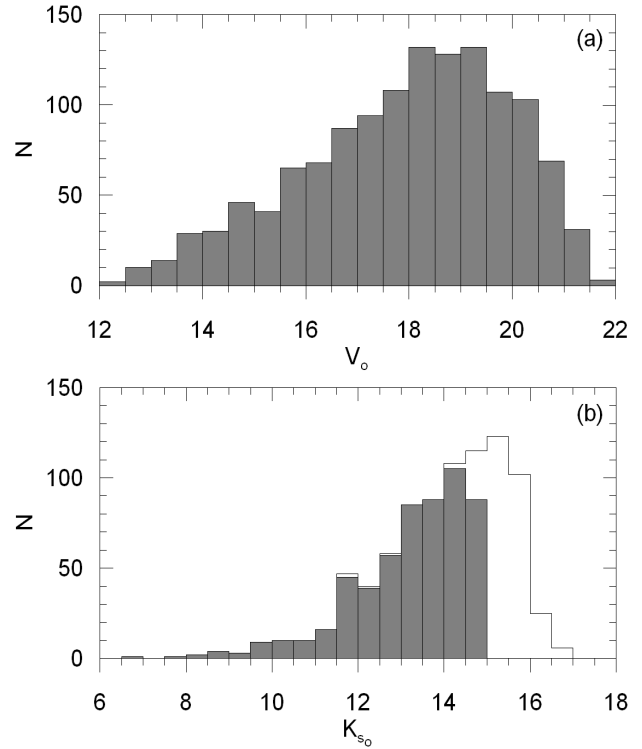}
\caption[] {Limiting magnitude of completeness for $V_o$ (a) and ${K_s}_o$ (b) magnitudes. The histogram with colours white and black corresponds to all stars (859) observed with $JHK_s$ photometry, whereas the one with black colour is plotted only for the stars with best quality, labeled with AAA (563).}
\end{center}
\end{figure}

\begin{figure}
\begin{center}
\includegraphics[scale=0.4, angle=0]{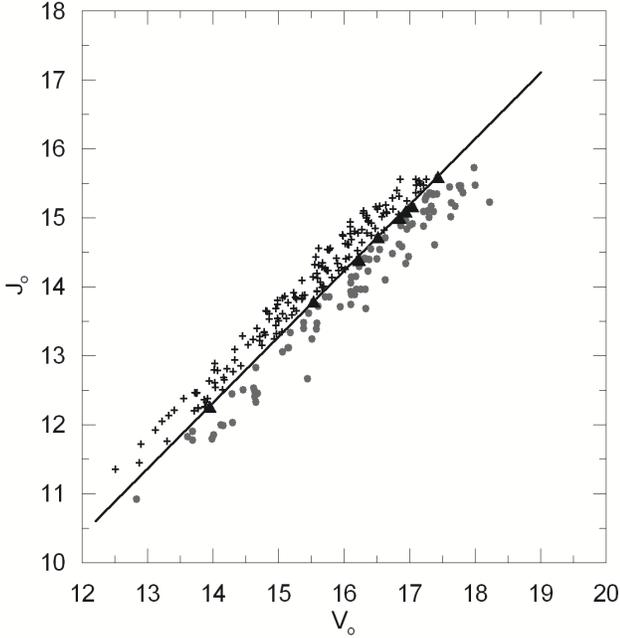}
\caption[] {Two magnitude diagram used for separation of dwarfs and giants (RC stars). Dwarf (plus), giant (circle), unclassified (triangle) due to their position on the diagram plus their errors. The solid line denotes the border of dwarfs and giants, adopted from \citet{Bilir06a}.}
\end{center}
\end{figure}

The RC stars in Fig. 10 were identified in two steps. First, we applied the colour constraint of \citet{Puzeras10}, i.e. $0.7\leq(B-V)_0\leq1.3$ and $0.3\leq(J-H)_0\leq0.6$ mag, and obtained a set of 249 dwarfs and giants. Then, we plotted them onto the $V_0-J_0$ two magnitude diagram (Fig. 13) and separated the giants (RC stars) from the dwarfs by the procedure of \citet{Bilir06a}. It turned out that 101 stars were classified as RC stars.                        

We evaluated the absolute magnitudes of 101 RC stars and transformed them to distances by the Pogson formula,
\begin{equation}
(m-M)_0=5\log d-5.  
\end{equation}
The $M_V$ absolute magnitudes for 90 stars with $BVI$ magnitudes were evaluated by the first equation in (8), whereas for $M_{K_s}$ absolute magnitudes for 11 stars with 2MASS magnitudes we used the last equation. The range of distance is $1<d\leq28$ kpc.

\subsection{Galactic Model Parameters Estimated with RC Stars}
\subsubsection{Density function}
We evaluated the logarithmic space densities $D^{*}=\log D+10$ for 101 RC stars in the distance range $1<r<24$ kpc, using the distances mentioned above. Here, $D=N/ \Delta V_{1,2}$; 
$\Delta V_{1,2}=(\pi/180)^{2}(A/3)(r_{2}^{3}-r_{1}^{3})$; $A$ denotes the size of the field (1.2 square-degrees); $r_{1}$ and $r_{2}$ are the lower and upper limiting distances of the volume $\Delta V_{1,2}$; $N$ is the number of stars per unit absolute magnitude; $r^{*}=[(r^{3}_{1}+r^{3}_{2})/2]^{1/3}$ is the centroid distance of the volume $\Delta V_{1,2}$; and $z^{*}=r^{*}\sin b$, $b$ being the Galactic latitude of the field centre. The results are given in Table 3.

\begin{table}
\centering
{\scriptsize
\caption{Space density function for 101 RC stars in the direction to the field SA 141, calculated by the distances evaluated via the absolute magnitudes based on the procedure in this study.}
\begin{tabular}{cccccc}
\hline 
    $r_1-r_2$ & $\Delta V_{1,2}$ & $r^{*}$    & $z^{*}$     & $N$     & $D^{*}$ \\ 
    (kpc)     & (pc$^3$)         & (kpc)     & (kpc)   &         &            \\
\hline   
 1-3  & 3.17E(6) &  2.41 &  2.40 &  9    & 4.45  \\
 3-5  & 1.19E(7) &  4.24 &  4.22 & 12    & 3.96  \\
 5-8  & 4.72E(7) &  6.83 &  6.81 & 14    & 3.56  \\
 8-12 & 1.48E(8) & 10.38 & 10.36 & 20    & 3.17  \\
12-16 & 2.89E(8) & 14.28 & 14.24 & 17    & 2.72  \\
16-20 & 4.76E(8) & 18.22 & 18.17 & 19    & 2.53  \\
20-24 & 7.10E(8) & 22.18 & 22.12 &  8    & 2.05  \\
$>$24 & ---      & ---   & ---   &  2    & ---   \\
\hline
\end{tabular}%
}
\end{table}%

\subsubsection{Density laws}
We adopted the density laws of Basel group \citep{Buser98, Buser99}. Disc structures are usually parameterized in the cylindrical coordinates using radial and vertical exponentials:
\begin{equation}
D_{i}(x,z)=n_{i}e^{-z/H_{i}}e^{-(x-R_{o})/h_{i}},
\end{equation}
where $z=r\sin b$ is the distance from Galactic plane, $x$ is the planar distance from the Galactic center, $R_{0}$ is the solar distance to the Galactic center \citep[8 kpc;][]{Reid93}, $H_{i}$ and $h_{i}$ are the scaleheight and scalelength, respectively, and $n_{i}$ is the normalized local density. The suffix $i$ takes the values 1 and 2, for the thin and thick discs.

The density law for spheroid component is parameterized in different forms. The most common is the \citet{deVaucouleurs48} spheroid used to describe the surface brightness profile of elliptical galaxies. 

\begin{equation}
D_{s}(R)=n_{s}\exp[10.093(1-(R/R_{o})^{1/4})]/(R/R_{o})^{7/8}.
\end{equation}
Here, $n_s$ is the normalized density at the solar radius, $R$ is the (uncorrected) galactocentric distance in spherical coordinates. $R$ has to be corrected for the axial ratio $\kappa=c/a$, 

\begin{equation}
R = [x^{2}+(z/\kappa)^2]^{1/2},
\end{equation}
where,
\begin{equation}
x = [R_{o}^{2}+r^{2}\cos^{2} b-2R_{0}r\cos b \cos l]^{1/2}, 
\end{equation}
with $l$ and $b$ being the Galactic longitude and latitude, respectively.

\begin{figure}
\begin{center}
\includegraphics[scale=0.4, angle=0]{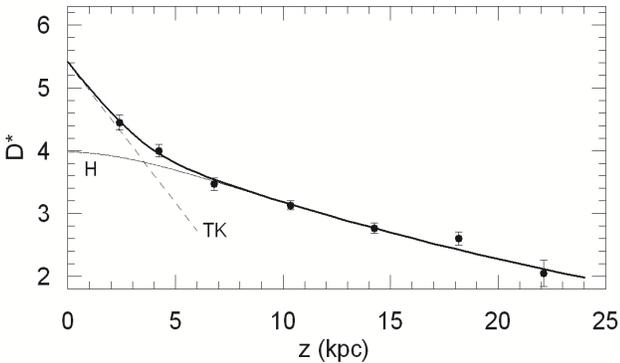}
\caption[] {Logarithmic space densities. The symbols TK and H denote the thick disc and halo components of the Galaxy, respectively. The curve indicates their combination.}
\end{center}
\end{figure}

\begin{table}
\setlength{\tabcolsep}{1.9pt}
{\scriptsize
  \centering
  \caption{Galactic model parameters estimated by the space density function in Table 3.}
    \begin{tabular}{crrrrrr}
\hline  
Thin Disc & \multicolumn{3}{c}{Thick Disc} & \multicolumn{3}{c}{Halo} \\
$n^{*}$    & \multicolumn{1}{c}{$n^{*}$} & \multicolumn{1}{c}{$\%$} & \multicolumn{1}{c}{$H$(pc)} & \multicolumn{1}{c}{$n^{*}$} & \multicolumn{1}{c}{$\%$} & \multicolumn{1}{c}{$\kappa$} \\
\hline    
6.61  & $5.40^{+0.05}_{-0.07}$& $5.80^{+0.7}_{-0.9}$ & $975^{+48}_{-48}$ & $4.02^{+0.07}_{-0.07}$ & $0.2^{+0.1}_{-0.1}$ & $0.89^{+0.11}_{-0.08}$ \\
\hline            
\end{tabular}
}
\end{table}

\subsubsection{Galactic model parameters}

We estimated the Galactic model parameters by fitting the density functions in Table 3 derived from the observations (combined from the three population components) to a corresponding combination of the adopted population-specific analytical density laws (Fig. 14). The distance to a star in the line of sight ($r$), in our sample, is rather close to its distance from the Galactic plane ($z$), due to the position of the field SA 141, i.e. $b=-85^{o}.83$. Hence, the space densities and the density laws are given as a function of $z$, instead of $r$. We extrapolated the logarithmic density from the nearest point, $z^{*}=2.4$ kpc, to the local space density for giants ($D^{*}=6.64$) in the literature \citep{Gliese69} and used the classical $\chi^{2}_{min}$ statistic to estimate Galactic model parameters, which is the most commonly used method in recent studies \citep{Du06, Juric08, Bilir08}. The results are given in Table 4. There is a good agreement between the Galactic model parameters estimated in this study and the ones appeared in the literature \citep[cf.][]{Karaali04, Cabrera05, Bilir06b, Bilir06c}.

\section{Discussion}

The RC stars are the most numerous population, within the red giants, and they are easy to detect at larger distances from the Sun. Hence, they become a powerful tool to trace the different Galactic components in the Milky Way. A lot of work has been done on this topic in recent years, but even more is still to come with the advent of the recent deeper databases in the NIR as those of UKIRT Infrared Deep Sky Survey \citep[UKIDSS;][]{Lawrence07}, or VISTA Variables in the Via Lactea (VVV) public survey \citep{Minniti10}, or even with high quality data at larger wavelengths as the ones that the more recent Wide Field Survey Explorer \citep[WISE;][]{{Wright10}} provides.

In recent years much work has been devoted to studying the suitability of RC stars for application as a distance indicator. The absolute magnitude of these stars in the K band is $M_{K_s}=-1.613$ mag with negligible dependence on metallicity \citep{Alves00, Laney12}. Whereas, their absolute magnitudes in the optical range lie from $M_V=+0.7$ mag for those of spectral type G8 III to $M_V=+1$ mag for type K2 III \citep{Keenan99}. 

In this work, we calibrated the $M_V$, $M_g$, $M_J$, and $M_{K_s}$ absolute magnitudes of RC stars in terms of colours. We found that the absolute magnitudes $M_V$ and $M_g$ are depend strongly on colour, whereas the $M_J$ and $M_{K_s}$ are weak. The calibrations of $M_V$ and $M_{K_s}$ absolute magnitudes are tested on 101 RC stars in the field SA 141. The Galactic model parameters estimated with this sample are in good agreement with the ones appeared in the literature. The data of the RC stars are taken from the {\em Hipparcos} catalogue and only stars with relative parallax errors $\sigma_{\pi}/\pi \leq 0.1$ are considered to obtain reliable absolute magnitudes and distances. The range of the metallicity of stars in the {\em Hipparcos} catalogue is narrow. Hence, we did not consider the metallicity effect on the calibration of absolute magnitude in terms of colour.

\section{Acknowledgments}
This work has been supported by the Scientific and Technological Research Council (T\"UB\.ITAK) 210T114.

We are grateful to Dr. Siegel for providing us the $UBVRI$ photometric data for the field SA 141 and Dr. T. G\"uver for reading and correcting for grammatic and linguistic aspect of the manuscript.

This research has made use of the NASA/IPAC Infrared Science Archive and 
Extragalactic Database (NED) which are operated by the Jet Propulsion Laboratory, 
California Institute of Technology, under contract with the National Aeronautics 
and Space Administration.

This publication makes use of data products from the Two Micron All
Sky Survey, which is a joint project of the University of
Massachusetts and the Infrared Processing and Analysis
Center/California Institute of Technology, funded by the National
Aeronautics and Space Administration and the National Science
Foundation. 

This research has made use of the SIMBAD, and NASA\rq s 
Astrophysics Data System Bibliographic Services.

\end{document}